\documentclass[journal]{vgtc}                
\ifpdf
  \pdfoutput=1\relax                   
  \usepackage{graphicx}                
  \DeclareGraphicsExtensions{.pdf,.png,.jpg,.jpeg} 
\else
  \usepackage{graphicx}                
  \DeclareGraphicsExtensions{.eps}     
\fi%

\graphicspath{{figures/}{pictures/}{images/}{./}} 

\usepackage{microtype}                 
\PassOptionsToPackage{warn}{textcomp}  
\usepackage{textcomp}                  
\usepackage{mathptmx}                  
\usepackage{times}                     
\usepackage{cite}                      
\usepackage{tabu}                      
\usepackage{booktabs}                  

\onlineid{1011}

\vgtccategory{Research}
\vgtcpapertype{algorithm/technique}

\title{A Domain-Oblivious Approach for Learning Concise Representations of Filtered Topological Spaces for Clustering}

\author{Yu Qin, Brittany Terese Fasy, Carola Wenk, and Brian Summa}
\authorfooter{
\item
 Yu Qin is with Tulane University. E-mail: yqin2@tulane.edu.
\item
 Brittany Terese Fasy is with Montana State University. E-mail: brittany.fasy@montana.edu.
\item
 Carola Wenk is with Tulane University. E-mail: cwenk@tulane.edu.
\item
 Brian Summa is with Tulane University. E-mail: bsumma@tulane.edu.
}

\shortauthortitle{Qin \MakeLowercase{\textit{et al.}}: A Domain-Oblivious Approach for Learning Concise Representations of Filtered Topological Spaces for Clustering}

\abstract{TBD%
} 


\teaser{
  \centering
  \includegraphics[width=\linewidth]{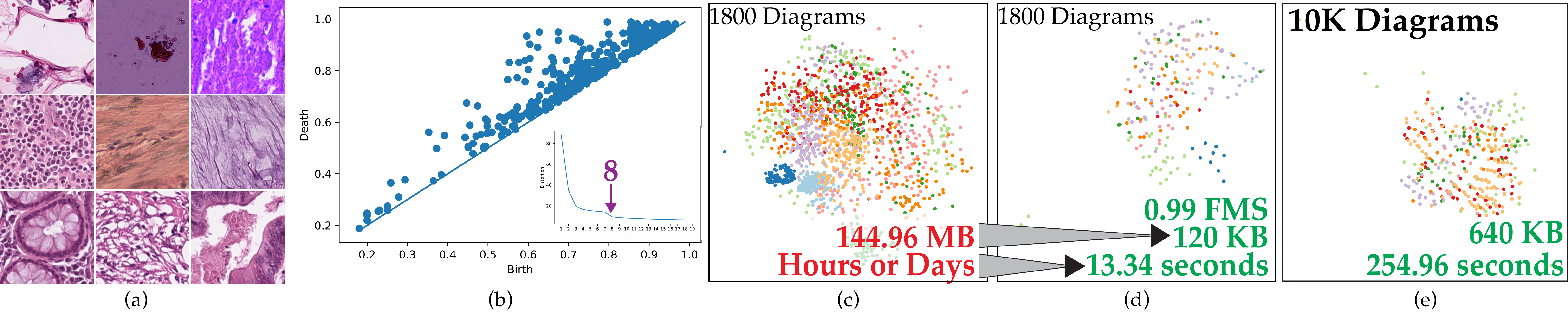}
  \caption{(a) Examples from a dataset of 10k histology images of colorectal cancer. (b) An example persistence diagram that encodes the topological structure in an image. The inset illustrates an elbow method plot run from clustering a subset of 1800 images using one-Wasserstein distance.  This shows there are approximately
  8 topologically distinct clusters. (c) Clustering result using one-Wasserstein distance on the subset. (d) Our high-quality concise representation uses only a fraction of the memory and computation time. (e) Our approach scales to the full dataset, which is not feasible with 1-Wasserstein.}
  \label{fig:teaser}
}

\abstract{Persistence diagrams have been widely used to quantify the underlying features of filtered topological spaces in data visualization. In many applications, computing distances between diagrams is essential; however, computing these distances has been challenging due to the computational cost. In this paper, we propose a persistence diagram hashing framework that learns a binary code representation of persistence diagrams, which allows for fast computation of distances. This framework is built upon a generative adversarial network (GAN) with a diagram distance loss function to steer the learning process. Instead of using standard representations, we hash diagrams into binary codes, which have natural advantages in large-scale tasks. The training of this model is domain-oblivious in that it can be computed purely from synthetic, randomly created diagrams. As a consequence, our proposed method is directly applicable to various datasets without the need for retraining the model. These binary codes, when compared using fast Hamming distance, better maintain topological similarity properties between datasets than other vectorized representations. To evaluate this method, we apply our framework to the problem of diagram clustering and we compare the quality and performance of our approach to the state-of-the-art.  In addition, we show the scalability of our approach on a dataset with 10k persistence diagrams, which is not possible with current techniques. Moreover, our experimental results demonstrate that our method is significantly faster with the potential of less memory usage, while retaining comparable or better quality comparisons.
}

\vgtcinsertpkg

\usepackage{cool}

\begin{document}


\firstsection{Introduction}

\maketitle

The features quantified by topological data analysis
(TDA)~\cite{edelsbrunner2010computational} have been shown to express the
fundamental structure of scalar fields in a way that is generally applicable to
many domains. TDA approaches---such as persistent
homology~\cite{edelsbrunner2000topological}, contour
trees~\cite{carr2003computing}, Reeb
graphs~\cite{biasotti2008reeb,pascucci2007robust}, and Morse(--Smale)
complexes~\cite{de2015morse,gyulassy2008practical}---have demonstrated ability
to extract
meaningful structure in a variety of research applications, including 3D shape
matching~\cite{carriere2015stable}, combustion
physics~\cite{bremer2010interactive,gyulassy2014stability}, nuclear
physics~\cite{maljovec2016topology}, fluid dynamics~\cite{kasten2011two},
chemistry~\cite{bhatia2018topoms,gunther2014characterizing}, Alzheimer’s
disease~\cite{lee2014hole}, autism spectrum
disorders~\cite{lee2011discriminative,shnier2019persistent}, cancer
histology~\cite{lawson2019persistent}, protein folding~\cite{xia2014persistent},
and bio-molecular analysis~\cite{meng2020weighted}. More generally, recent
work includes using TDA quantification as input to machine
learning~\cite{carriere2017sliced,reininghaus2015stable,rieck2019topological,lawson2019persistent}.

As described in detail in \secref{TDA_background}, a persistence diagram is a common way to present the topological structure in a dataset. The distance between these diagrams is often used to measure the topological (dis)similarity between data, which has important applications in scientific visualization~\cite{yan2021scalar}. Moreover, for approaches that cluster based on topological similarly\cite{carriere2015stable,lawson2019persistent,vidal2019progressive,turner2014frechet}, computing the distance between diagrams is a fundamental operation. However, computing these distances is costly in practice.

The most widely accepted
persistence diagram distance measures, the Wasserstein distances, require expensive matching
of all persistence points between two diagrams. As
discussed in \secref{related}, to combat this complexity,
many approaches have attempted to reduce this cost. The goal of our
work is the same, but provides significant advances over the state-of-the-art.
As we detail in this paper, we provide a new representation for expressing
topological structure that is more concise than previous works, but
also leads directly to faster computations of distances. In particular, we
show how to reduce diagrams to simple 64-bit binary codes. The key
to this representation is a learned hash function. As we show, this hash
can be learned purely from random, synthetically generated diagrams. We have
found that the only constraint on generating training data is that the generated
diagrams should have approximately the same average number of persistence points as
are in the test data. In other words, the training is domain-oblivious with a model
being potentially used on a wide variety of datasets without the need for
retraining. In this new representation, distances are calculated by a simple
bit-wise count comparison between binary codes (the Hamming distance). This makes
distance computation extremely fast and scalable. We illustrate this
scaling through the clustering of a dataset with 10k diagrams, a size which is
not achievable for several existing approaches.

\subsection{Contributions}
The specific contributions of this work are:
\begin{itemize}
    \vspace{-4pt}
    \item A concise binary code representation of persistence diagrams that maintains topological (dis)similarity in Hamming space;
    \vspace{-4pt}
    \item A procedure to train the binary code hash function that can run purely on synthetic data and therefore is domain-oblivious; and
    \vspace{-4pt}
    \item Applications to topological clustering of
        real-world datasets that provide: Significant comparison speedups, potentially lower
        memory footprints, and comparable or better quality clustering results
        than other vectorized representations of persistence~diagrams.
\end{itemize}

\section{Background and Related Work}
\label{sec:related}
This section outlines the technical background for persistent homology and
hashing, as relevant to the methods developed in this paper.

\subsection{Persistent Homology and Persistence Diagrams}
\label{sec:TDA_background}

Given a dataset, we view it as a topological space or a sequence of
(nested) topological spaces, called a \emph{filtration}. Then, we employ
homology and persistent homology, respectively,
to qualitatively and quantitatively describe it.
Homology is a concept from algebraic topology that captures the fundamental
structure of a topological space~\cite{munkres:at}. The structure is
qualitatively described
through the homology groups, whose generators we call \emph{features}. Each
feature has a dimension associated to it;
dimension zero features are connected components, dimension one features are
loops or tunnels, dimension two features in $\R^3$ correspond to voids, etc.
Persistent homology quantifies the homology of an entire
filtration~\cite{edelsbrunner2010computational}. In
particular, each feature~$f_i$ also has a birth
time~$b_i$ and a death time $d_i$, indicating the parameters of the filtration
for which that feature ``lives''; we call the difference~$d_i-b_i$ the
\emph{lifetime} or \emph{persistence} of the feature. We represent this feature
as the persistence point $(b_i,d_i) \in \R^2$.
Since a feature must
be born before it dies with respect to the filtration parameter, persistence
points are restricted to lying above the
\emph{diagonal} defined by the line $x=y$.\footnote{In general, it is possible
for some features to
 have a birth but no associated death.
 In our experiments,
these features are not as informative as the features with defined birth and
death times. For this reason, we do not include them in our definition above.}
The \emph{persistence diagram} is the
multi-set of birth-death pairs.

This abstract concept is best illustrated by describing example filtrations. For
scalar fields, a common filtration is the evolution of sublevel sets. The sublevel set filtration is the progression of a watershed transformation~\cite{beucher1979international},
where water sources grow from local minima (i.e., basins) of the field. The zero-dimensional features (i.e.,
connected components) are the watersheds that begin at
the local minima.
One-dimensional features form when a watershed completely surrounds a local
maximum (i.e., peak).
The lifetimes of these features
are recorded as the scalar value (i.e., water height) from where a feature first appears ({\em birth}) to
the value at which it merges with an ``older'' feature ({\em
death}). All birth and death
events occur at critical points.

For unstructured points (i.e., point cloud data that is given as a discrete set
of points with a pairwise distance defined), a filtration can be built from the evolution of a
Vietoris--Rips complex~\cite{vietoris1927hoheren,hausmann1995vietoris}.
Here, simplices are formed in the complex from an ever-increasing
neighborhood around each point. In this filtration, keeping track of the
connected components can detect the size and the number of distinct clusters and
recording the evolution of one-dimensional features can detect the presence and size of
circular features in the data. These features are agnostic to the domain of
the data or even the dimension of the space. For this reason, one can say that these
features represent the fundamental structure of
data.

\begin{figure}
    \centering
    \includegraphics[width=\columnwidth]{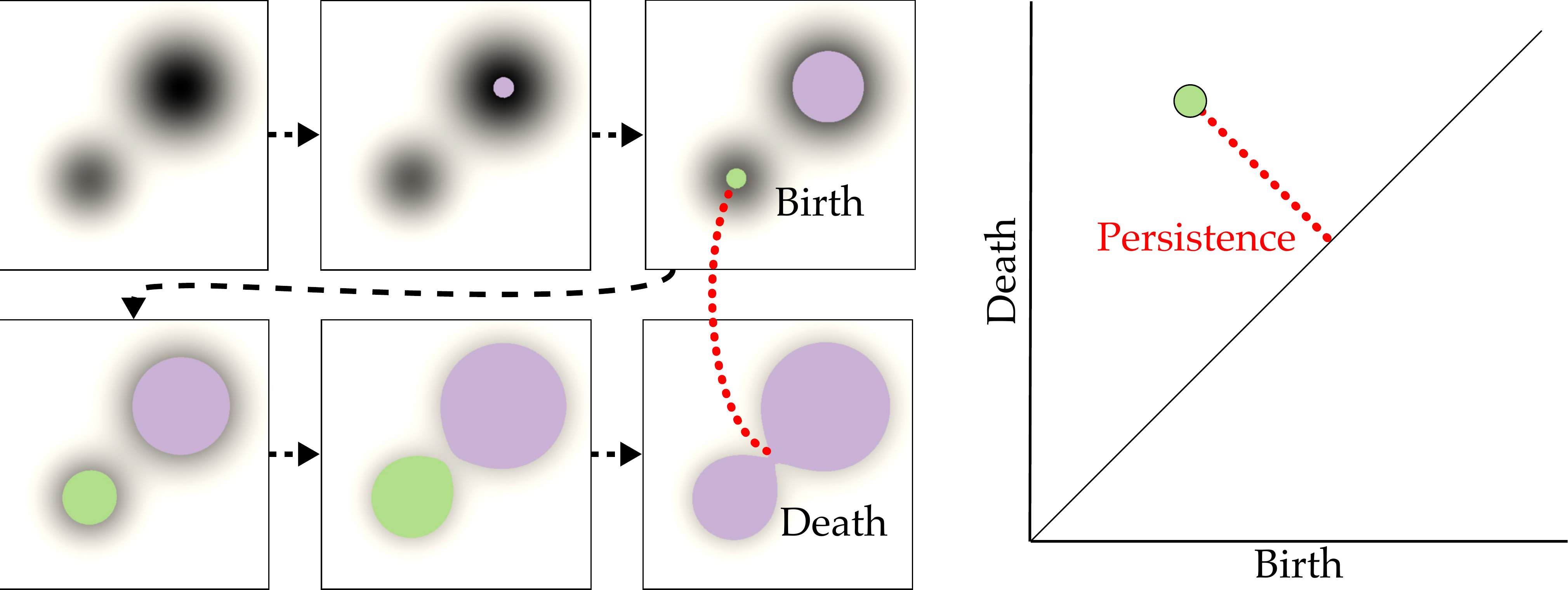}
        \caption{Progression of a sub-level set of a scalar field.  The feature
        (green) is born at a minimum and dies when it merges with an older
        feature (purple). The birth and death are represented as a point in the
        zero-dimensional persistence diagram.  The $L_{1}$ distance from this point to
        the diagonal is the lifetime, or persistence, of the feature (red).
        }\label{fig:persistence}
        \vspace{-14pt}
\end{figure}

\figref{persistence} illustrates a simple example of the zero-dimensional features
of a sublevel set filtration on a scalar field with two basins. As the sublevel
set increases, first the purple feature is born in the deepest basin, followed
by the green feature. As the filtration continues, the green feature eventually
dies when it merges with the purple connected component. The green feature is
represented in the diagram by plotting these birth and death values. Note, that
since the purple feature has not yet died, it is not yet in the diagram.
%

\paragraph{Wasserstein and Bottleneck Distances}
Letting $\dgmspace$ denote the collection of all diagrams,
the $q$-Wasserstein distance
$d_q \colon \dgmspace \times \dgmspace \to \R$ is defined by
$$
    d_q(p_1,p_2) := \min_{M} \left( \sum_{(a,b)\in M} ||a-b||_{\infty}^q +
    \frac{1}{2^{q-1}} \sum_{a \in M^c} |a_x-a_y|^q
\right)^{1/q},
$$
where $M$ ranges over all matchings between persistence diagrams $p_1$ and $p_2$, and $M^c$ is the
set of persistence points in $p_1 \sqcup p_2$ that do not appear in the matching
$M$; see~\cite{kantorovich2006translocation,monge1781memoire,kerber2017geometry,cohen2010lipschitz}.
The Wasserstein distance optimizes a matching between two diagrams and sums the distances between matched points ($M$) as well as the point-to-diagonal distances for unmatched points ($M^c$).
Letting~$q\to \infty$, gives the bottleneck distance (or, interleaving distance)~\cite{cohen2007stability}.
Setting~$q= 1$ gives the one-Wasserstein distance (W1), a popular choice in
applications and therefore the target of our
work\cite{carriere2017sliced,adams2017persistence}. Note that when we refer to
the Wasserstein distance without specifying $q$, we are referring to~$q=1$.
These diagrams are (Lipschitz) stable in the
presence of slight perturbations or noise in data~\cite{cohen2010lipschitz}.
Given this
stability, diagrams that are close in distance are often considered to be
topologically similar. 
Computationally, however, both the Wasserstein and the bottleneck distances are
expensive, as they require computing the optimal matching between
persistence points in the two diagrams. In particular, computing the Wasserstein distance
between two diagrams takes $O(n^2\log^2 n)$ time, where~$n$ is
the total number of simplices (which, in turn, can be exponential in the size
of the input data)~\cite{vaidya1989geometry}.

\paragraph{Approximating Distances}
When many distances between diagrams need to be computed, the roughly quadratic
computation can be daunting.
Thus, several approaches
have been introduced to approximate computing the Wasserstein
distances~\cite{kerber2017geometry,
bertsekas1981new,agarwal2000vertical,
soler2018lifted,sheehy2021sketching}.
One approach that is quite successful is to simplify the input representation,
before a persistence diagram is even
computed~\cite{wagner2012efficient,henselman2016matroid}.
However, this makes assumptions on the underlying domain (e.g., a 2D
or 3D image).

Kerber et al.~\cite{kerber2017geometry}
introduced an approximate Wasserstein distance algorithm to accelerate the
computation of the
matching using a $k$-d tree. This iterative computation bounds either
quality or time to approximate the Wasserstein distances; we call this
distance the progressive Wasserstein (\PW) distance.
This algorithm was extended by Vidal et
al.~\cite{vidal2019progressive} for the problem of computing barycenters of
persistence diagrams.
Although very fast, these approaches still require the pairwise matching and, as
we illustrate in \secref{performance}, the memory requirements can be significant as data
sizes grow.

\begin{figure*}[ht]
    \centering 
    \includegraphics[width=\linewidth]{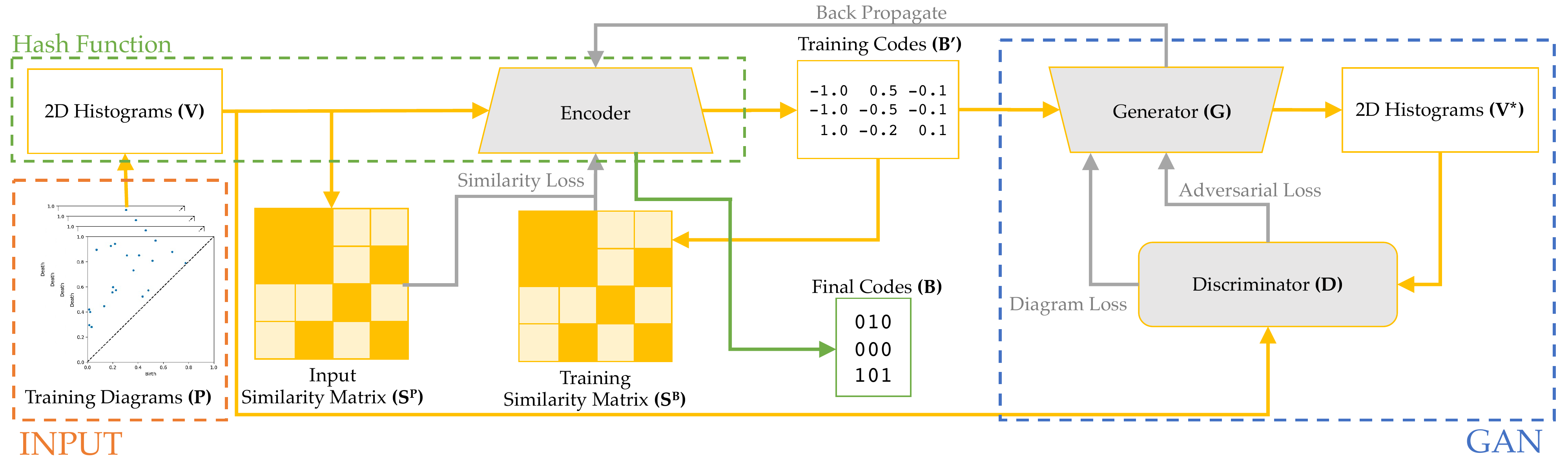}
    \vspace{-18pt}
    \caption{Training architecture of the PD-GAN model that takes a set of
    persistence diagrams~$\dgmset$ as input and learns corresponding binary
    codes~$\mathbf{B}$.
    Yellow arrows depict flow of diagrams through training, while grey arrows represent information
    flow (loss functions, etc.). A Generative adversarial network (GAN)
    is used to train the encoder of binary codes such that the similarity matrix of codes, $\mathbf{S}^B$, closely matches the similarity of the input diagrams, $\mathbf{S}^P$. Our final hash function that takes in a diagram and produces a binary code is highlighted in green.
    }
    \label{fig:PD-GAN}
    \vspace{-14pt}
\end{figure*}

In many circumstances, the diagrams we have are bounded, that is, there exists a
square $D \subset \R^2$ such that all persistence points lie inside $D$.
Then, for a given input parameter $d \in \N$,
we create a $d \times d$ grid
over~$D$. Using this grid, we define a histogram, where we count the number of
persistence points in each grid cell. Fasy et al.~\cite{fasy2018approximate} use these histograms to design a data
structure that supports searching for near neighbors based on the bottleneck
distance and Lacombe et al.~\cite{lacombe2018large} computes
Wasserstein distances (optimal transport in this space) between the histograms.
While this work can benefit from several fast optimal transport computation
approaches~\cite{cuturi2013sinkhorn, solomon2015convolutional}, it still poses
significant costs for distance computations.
We call the Wasserstein distance between
histograms the Histogram Wasserstein~(\OT) distance.

\subsection{Other Topological Descriptors}
\label{sec:related_distances}
The previous approaches for histograms
can be thought of as representations of the distribution of the persistence
points. In this line of research, density estimators built over the persistence points
have been studied by several groups of researchers~\cite{adams2017persistence,anirudh2016riemannian,reininghaus2015stable,rieck2019topological,chen2015statistical}.
One benefit of considering density estimators
is that they can be vectorized.
With vectorized representations, the space cost of the discretization can be
weighed against the speedup gained by computing $L_p$ distances between vectors
instead of distances between persistence diagrams.

\paragraph{Persistence Images (PI)}
The persistence image (\PI) is a discretization of
a weighted kernel density estimator (a non-parametric density estimator) built on the rotated points of a persistence diagram~\cite{adams2017persistence}). Roughly, these images estimate the density of points by summing a Gaussian
kernel centered at each point.
In practice, using the \PI
requires choosing: a bounding box, a discretization resolution, and a weight
function on the set of points in the persistence diagram.
Choosing these parameters can be non-trivial
in practice, but various heuristics have been successfully
employed~\cite{zhao2019learning,chen2015statistical,adams2017persistence}.

\paragraph{Betti Curves (BC)}
Other topological invariants include the Euler characteristic and the Betti numbers. For
a given integer $k \in \N$, the
$k^{th}$ Betti curve
(BC)~\cite{robins2002computational,gameiro2004topological,xu2019finding} is the
rank of the dimension $k$ homology group with respect to the filtration
parameter. Roughly, this is the count of the topological features present as the
filtration parameter increases. Each feature associated to a persistence point
$(b_i,d_i)$ contributes $+1$ to the Betti curve in the interval from $b_i$
to~$d_i$. Hence, the more persistent a feature, the more it contributes to the Betti
curve.
The $L_p$ distance between Betti curves can be computed explicitly, or can be
approximated by sampling or discretizing the domain. Often, the latter approach is preferred in
practice.
Hence, using the Betti curves requires selecting the dimension(s) of
interest, a bounding box, and a discretization resolution.

\subsection{Learning to Hash}
Given the ability of binary codes to significantly boost distance computation for searching, hashing
methods have attracted increasing attention for large-scale approximate
nearest-neighbor search
\cite{andoni2006near,kulis2009learning,kulis2009kernelized,raginsky2009locality}.
In this paper, we focus on using unsupervised machine learning to build a good hash function that maps high-dimensional data into low-dimensional
Hamming space.

Unsupervised building of hash functions can be roughly
divided into two groups: non-deep hashing and deep hashing. Typical non-deep
hashing includes PCA hashing~\cite{wang2012semi},
spectral hashing~\cite{weiss2009spectral}, and iterative
quantization~\cite{gong2012iterative}, which all attempt to preserve a pairwise similarity of the original data in their resulting binary codes.
For example, spectral hashing uses
eigenfunctions of the data similarity graph to build their hash.
More recently, deep hashing
\cite{gao2017video,do2016learning,lin2016learning,song2018deep} has been
introduced due to the great advances made in deep learning. The non-linear structure
of a convolutional neural network (CNN) can extract multiple
hierarchical feature representations of input data and learn their nonlinear
relationships to build a binary representation. However, the need for data to be
labeled for CNNs means that unsupervised approaches to learn a hash function cannot take full advantage of a deep
learning model. Inspired by the introduction of the generative
adversarial network (GAN)~\cite{goodfellow2014generative}, other work focuses
on the unsupervised learning of hash functions using a GAN without the need for
labeled data~\cite{cao2018hashgan,song2018binary}.
Overall, previous hashing approaches mainly focused on the image retrieval tasks, which
live in Hilbert space and have nice statistical properties.
In our work, we show that a natural image hashing approach~\cite{song2018binary} can be used to hash persistence diagrams in non-Hilbert space. In doing so, we present the first approach to transform topological features into a binary
representation.

\section{Learning Topological Binary Codes}
\label{sec:methods}
In this work, we explore the use of concise binary codes to represent
persistence diagrams.
Below,  our process takes as input $N$ persistence
diagrams~$\dgmset=\{p_i\}_{i=1}^N$ and trains neural networks to hash
persistence diagrams to binary codes.  Then, the set of binary codes and their
Hamming distances can be used in lieu of persistence diagrams and their
Wasserstein distances.
See
\figref{PD-GAN} for an illustration of the architecture for our approach.

\subsection{Vectorizing Input}
\label{sec:vectorization}
The first step in our hash function is to take as input a set of
diagrams~$\dgmset=\{p_i\}_{i=1}^N$ and to convert it into a vectorized form appropriate
to use as input to train networks.  As mentioned in \secref{related},
we have several choices for vectorized representations of the input persistence
diagrams, including 2D histograms using Wasserstein distance  (\OT)~\cite{lacombe2018large},
persistence images using $L_2$ distance~(PI)~\cite{adams2017persistence},
and Betti curves using $L_2$ distance~(BC)~\cite{gameiro2004topological}.
Using the parameters outlined in the respective papers, we compare the use of
these three vectorizations for clustering relative to clustering persistence
diagrams using Wasserstein distance.
\tabref{fm-vectorized} provides the results
using the Fowlkes-Mallows score (FMS), the evaluation measure used in this work (see \secref{evaluation}).
The scores can be in the interval~$[0,1]$, with a score of 1 indicating a perfect match.
As \tabref{fm-vectorized} illustrates, \OT provides the
most accurate clustering. Therefore, we use 2D histogram vectorization for our training and hashing.

\begin{figure}[thbp]
  \centering
    \includegraphics[width=0.75\columnwidth]{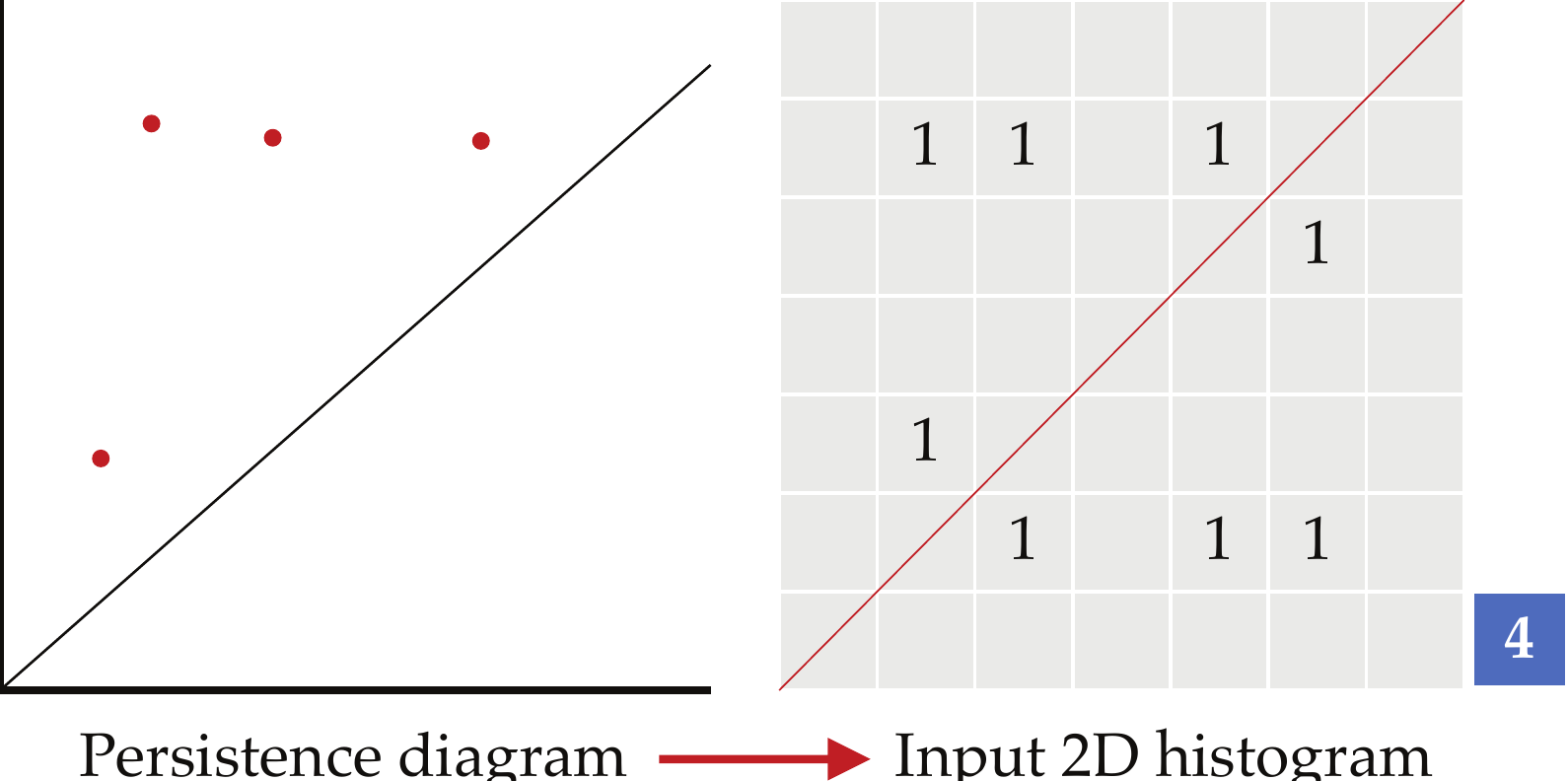}
  \caption{The initial 2D histogram representation used for training and an intermediate step before hashing.}
  \label{fig:2DHash}
\end{figure}

We transform our diagrams to
histograms on a 2D uniform grid of size~$50\times 50$ on~$[0,1]^2$ with the entropic term: $0.1/avg_{i=1}^N|p_i|$ (the recommended parameter values \cite{lacombe2018large}). Each cell of
the grid counts the number of persistence points in the diagram that lie inside each, with an
additional cell that contains a count of the total number of
persistence points. In addition, similar to Reininghaus et al.~\cite{reininghaus2015stable}, we augment this representation by reflecting counts across
the diagonal to the empty space below, see \figref{2DHash}.
We found that this
augmentation improves the quality of our clustering results over the standard
histogram (improvement of $3\%$ for the 3D Shape-1 dataset of \secref{datasets}).
In summary, the first step of training our hash function is to
convert each
diagram~$p_i \in \dgmset=\{p_i\}_{i=1}^N$ into a 2D histogram~$v_i$.  In
\figref{PD-GAN}, $\histset$ denotes the set of all such histograms.
\camera{I edited above for flow, but I think we should present the
vectorization here, and move the justification (table 1 \& it's discussion) to
an appendix.  Forward references can be frustrating to a reviewer, and a few
things are not yet defined here (namely, FMS and 3D Shape-1.}
\begin{table}[tbh]
  \caption{Comparison of vectorized representations using Fowlkes-Mallows score. Values closer to one are better.}
  \label{tab:fm-vectorized}
  \scriptsize%
	\centering%
  \begin{tabu}{%
	lrrrr%
	*{7}{c}%
	*{2}{r}%
	}
  \toprule
 Dataset &HW &PI &BC \\
  \hline
  	3D Shape-1 &\textbf{0.98}  &0.81  &0.8  \\
  \bottomrule
  \end{tabu}%
\end{table}

\subsection{Similarity Matrix}
\label{sec:similarity_matrix}
Before training our model, we need one more object: a $N\times N$
similarity matrix,~$\mathbf{S}^P=\{s_{ij}^P\}_{i,j=1}^N$,
\camera{switch $^P$ to $^V$ since we are referring to histogram similarity not
persistence diagram similartiy}
that stores all pairwise similarities between
histograms in~$\histset$.   We explore two different methods to define similarity. Our first
approach is to invert a topological distance measure. The most natural choice
is to invert Wasserstein distance, since this is the
distance that we would like to mimic using Hamming distances of binary codes.  As our
experiments in \secref{results} illustrate, computing these distances for potentially
thousands of training datasets is prohibitively expensive.  Therefore we adopt
the HW distance on~$\histset$ as used in Lacombe et al.~\cite{lacombe2018large}. This gives us a computationally feasible way to build the matrix, and also aligns well with our choice of intermediate vectorization. Specifically,
we compute a real-valued similarity matrix with~$s_{ij}^P := 1 - d_1(v_i,v_j)$.
\camera{present only the similarity that we use here. then, justify the choice
either in experiments or in an appendix}

To give more flexibility to the learning algorithm, we also explored the use of
a less rigid binary similarity matrix.  Rather than using the real/original values above,
the matrix is formed by setting $s_{ij}^P=1$ if $v_i$
and $v_j$ are similar and $s_{ij}^P=-1$ if they are dissimilar.  Defining this similarity
can take many forms.  We found that thresholding based on the closest
k-nearest neighbors for each diagram was not sufficient. This had the
tendency to count distant persistence diagrams as similar if a training diagram was not
close to many others (also, close diagrams were treated as dissimilar if a
diagram had many neighbors).  We opted to use a rejection approach where
all diagrams that have distance greater than the mean distance value are
rejected as dissimilar. This is done in two passes on a per row (diagram) basis.
This approach allows the binary labeling to have a soft threshold that maintains
small distances, discounts large distances, and is less rigid than a nearest
neighbor approach. As we show in \secref{results}, this flexible
binary matrix approach outperforms real-valued distances.

\subsection{PD-GAN Model}
\label{sec:GAN}
We now describe how we generate a 64-bit binary code by learning a hash
function $h \colon \dgmspace \to \{0,1\}^{64}$, where $\dgmspace$ is the original space of diagrams.
The hash is designed to maintain topological
similarity when comparing binary codes with Hamming distance. Our learning framework uses the image hashing approach of Song et
al.~\cite{song2018binary}.
Below we describe their approach and how it is used in our context.
To build our hash function, two key parts are trained for the PD-GAN model, the
encoder and the GAN; see \figref{PD-GAN}:

\paragraph{Encoder}
The encoder extracts the features of input diagrams based on a pretrained VGG19
network \cite{szegedy2015going,song2018binary} with five groups of convolution
layers with max pooling. The number of filters in each of these groups are 64,~128,~256, and
512. The output size of the last fully connected layer is the bit length $L=64$.
The training of the encoder is driven by minimizing a {\em similarity loss}
function; see \parref{similarity}.\camera{this is a forward reference and
confusing to follow ...} After training, each binary code,
$\mathbf{h}(v_i) = b_i \in \mathbf{B}$, can be formed from the signs of the
values of the last fully connected layer when $v_i$ is run through the encoder
(i.e.,~$\mathbf{h}(v_i)^{k} = sgn(x^k$), where $x^k$ is the $k$-th value in the last layer). We call this last layer the binary-code layer. However, a non-smooth, binary representation can be problematic for the gradient computation needed in training. To avoid this problem, an intermediary, real representation of the binary code, $\binarycodeset'=\{\mathbf{h'}(v_i)\}_{i=1}^N$, is used during training:
$$
\mathbf{h}'(v_i)^k = \left\{ \begin{array}{rcl} +1 & \mbox{for} & x^k \geq 1 \\ x^k  &
\mbox{for} & 1\geq x^k \geq-1 \\ -1 & \mbox{for} & x^k \leq -1,
\end{array}\right.
$$
where $k$ is the $k$-th element of the binary code.\camera{BTF:I am now confused
about what $k$ is ...} After training, the binary
codes, $\mathbf{B}$, can be extracted from the binary-code layer as described~above.


\paragraph{GAN: Generator and Discriminator}
To improve the accuracy of the learned hash function, a GAN~\cite{goodfellow2014generative,song2018binary} is used. See \figref{PD-GAN} (blue). Specifically, the
generator $\mathbf{G}$ can be considered as an inverse encoder,
where the output of the encoder is used as the input to the network with four
deconvolutional layers. $\mathbf{G}$ creates a set of synthetic
histograms,~$\mathbf{V}^{*}$, from their training codes, $\mathbf{B}'$. The
generator's goal is to create a~$\mathbf{V}^{*}$ which cannot be distinguished
from $\mathbf{V}$ by the discriminator $\mathbf{D}$. $\mathbf{G}$ is trained by
minimizing {\em diagram loss} and {\em adversarial loss} functions defined
below. The discriminator informs the generator to improve~$\mathbf{V}^{*}$, while the generator then informs the encoder to improve the hash function.

\subsubsection{Loss Functions}
\label{sec:optimization}
Loss functions need to be defined for the above components to minimize: similarity loss, diagram loss, and adversarial loss.

\paragraph{Similarity Loss} \label{p:similarity}
Given the  similarity matrix $\mathbf{S}^P=\{s^P_{ij}\}_{i,j=1}^N$ from
\secref{similarity_matrix} and the training codes $\binarycodeset'$,
the similarity loss captures a direct
connection between our binary representations and topological distances.
%
%
Let $\mathbf{S}^B=\{s^B_{ij}\}_{i,j=1}^N$, where
$s^B_{ij} = \frac{1}{L}\mathbf{h'}(v_i)^T\mathbf{h'}(v_j)$ and $L$ is the bit length.
%
Then the similarity loss is defined as:


$$
l_{sim} =
\frac{1}{2} ||\mathbf{S}^B - \mathbf{S}^P||^2 + ||\mathbf{B'} - \mathbf{B}||^2\;,
$$
where $||.||$ are Euclidean norms.
%

\paragraph{Diagram Loss} Intuitively, the diagram loss compares the
generated histograms,
$\mathbf{V}^{*}$, to the corresponding input histograms, $\mathbf{V}$. The diagram loss function is the combination of a pixel-wise Mean Squared Error (MSE), and the perceptual loss~\cite{song2018binary}. Perceptual loss is given by the last layer of the discriminator, $\mathbf{D}$.
Perceptual loss accounts for the observation~\cite{ledig2017photo} that
 pixel-wise MSE optimization often lacks high-frequency content. We verified experimentally in our test dataset that including perceptual loss provided more accurate results.  These two losses are summed to form the diagram loss: $l_{dia} = l_{mse} + l_{perceptual}$.

\paragraph{Adversarial Loss} The adversarial loss is designed to improve the reconstruction quality of
generator $\mathbf{G}$ and
is defined as
$l_{adv} = \log (\mathbf{D}(\mathbf{V})) + \log
(1-\mathbf{D}(\mathbf{V}^{*}))$.

\paragraph{Combined Loss} Finally, the combined loss used in training is the weighted sum of the three
losses: $l = l_{sim} + {\omega}_1 l_{dia} +  {\omega}_2
l_{adv}$, where ${\omega}_1={\omega}_2=0.1$ were used in our experiments as in Song et al.~\cite{song2018binary}.

\subsubsection{Learning}
\label{sec:solving_optomization}
To train the PD-GAN, the loss functions are minimized in the following steps. First, the training codes $\mathbf{B'}$ are created by the encoder, using parameters $\phi$ for the VGG19 part of the encoder and $\mathbf{W}$ for the binary-code layer.
Then the generator $\mathbf{G}$ reconstructs the
diagrams $\mathbf{V}^{*}$ with a parameter vector $\theta$.
%
The discriminator $\mathbf{D}$ uses the parameter vector $\sigma$.
%
Back-propagation for learning and stochastic
gradient descent are used to find the (locally) optimal parameters based on the loss functions. Specifically, the parameters $\{\phi,\theta,\sigma,\mathbf{W}\}$ are
updated during each iteration, where $\tau=0.001$ is the default learning rate in our experiments:

\begin{align*}
    \mathbf{W} & \leftarrow \mathbf{W} - \tau \bigtriangledown_\mathbf{W}(l_{sim}+l_{dia})\\
    \phi & \leftarrow \phi - \tau \bigtriangledown_\phi(l_{sim}+l_{dia})\\
    \theta & \leftarrow \theta - \tau \bigtriangledown_\theta(l_{dia} + l_{adv})\\
    \sigma & \leftarrow \sigma + \tau \bigtriangledown_\sigma l_{adv}
\end{align*}

\begin{figure}[tb]
    \centering 
    \includegraphics[width=\columnwidth]{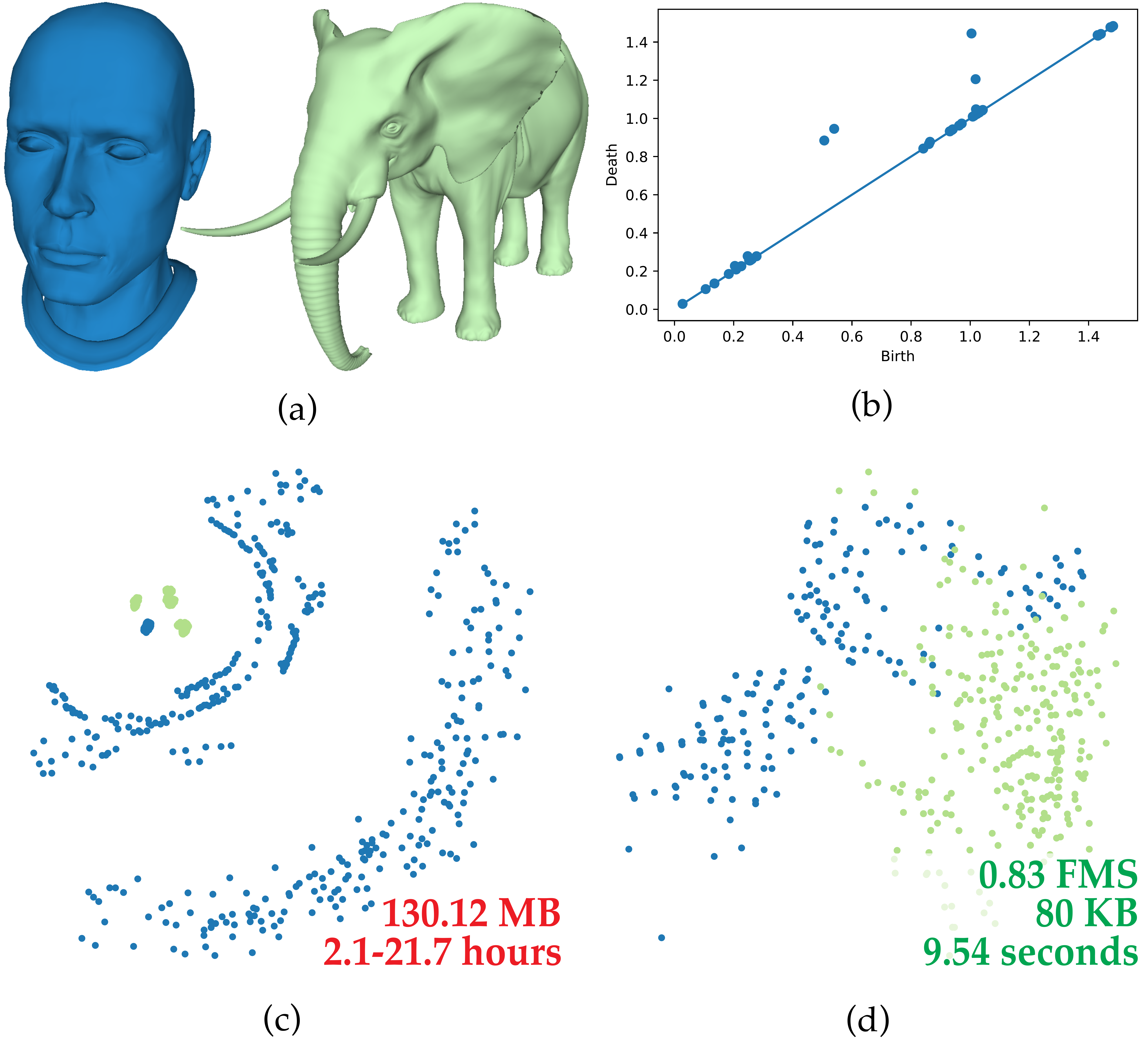}
    \caption{3D shape-1: The representative topological clusters are shown (a) by color with an example persistence diagram (b). Two MDS plots show results (c) based on Wasserstein distance and (d) Hamming distance for our generated binary codes. The FMS score is the clustering performance measure between two sets of clusters (c) and (d).}
    \label{fig:3DShape-1}
\end{figure}

\subsection{Hash Function and Distance Computation}
After training, diagrams can be hashed by first converting them into their 2D histogram representation and then running each through the PD-GAN encoder to map each to a 64-bit integer. See~\figref{PD-GAN} (green). A direct consequence of this binary encoding is that the representation is concise and distances between codes are computed in Hamming space.  Distance computations are now a simple bit-wise operation: a population count of the XOR of the bits ($popcount(X \mathbin{\oplus} Y)$ for binary codes $X$ and $Y$).  Not only is this distance computation simple, it is also supported in hardware on modern CPUs.  In \secref{results} we show the speed of this computation in our standard Python implementation, but also with a C++ implementation that leverages this hardware support.

\section{Experimental Results}
\label{sec:results}

To illustrate the effectiveness of our approach, we use our generated binary representation in clustering applications. We experimented on five datasets and refer to the dataset information in \secref{datasets}.
 In order to evaluate the quality of the distance approximations of our approach, we apply a distance-based single-linkage hierarchical clustering algorithm by using the scikit-learn \cite{scikit-learn} Python library. 
 It takes a {\em distance matrix} as input, which contains all pairwise distances between histograms.
 The objective of single-linkage hierarchical clustering is to produce a nested sequence of partitions by
successively merging clusters in a bottom-up fashion
until $k$ clusters in total are reached.

For the datasets that have undefined $k$ topological clusters, we use the elbow method \cite{thorndike1953belongs} to determine the number of clusters. 
%
For this we deploy the hierarchical clustering for a sequential set of potential values for $k$ and then plot the total within-cluster sum of square distances
versus $k$
(which measures the compactness of the clustering). The final number $k$ of clusters is then chosen by the elbow of the curve.

Results are evaluated against clustering results using the Wasserstein distance. Test datasets are described in \secref{datasets}.
 Evaluation methods are detailed in \secref{evaluation}.  Next, we describe how training can be domain-oblivious in \secref{domain}.  Finally, performance and quality results are reported in \secref{performance}.

\subsection{Datasets}
\label{sec:datasets}
We evaluated the clustering of five datasets that include 3D shapes, ensemble simulation data, and 2D medical images.

\textbf{3D Shape-1}\cite{sumner2004deformation}: This dataset contains 6
different 3D shape classes including camels, horses, elephants, cats, human heads, and
faces. We created 200 persistence diagrams for each class using the implementation of
Carri{\`e}re et al. \cite{carriere2015stable} to produce a Vietoris--Rips filtration. This previous work showed that there were $k=2$ distinct
topological clusters in this dataset. The average number of persistence points
per diagram is 63, ranging from 49 to 100. \figref{3DShape-1} shows example shapes and a persistence diagram.

\begin{figure}[tb]
    \centering 
    \includegraphics[width=\columnwidth]{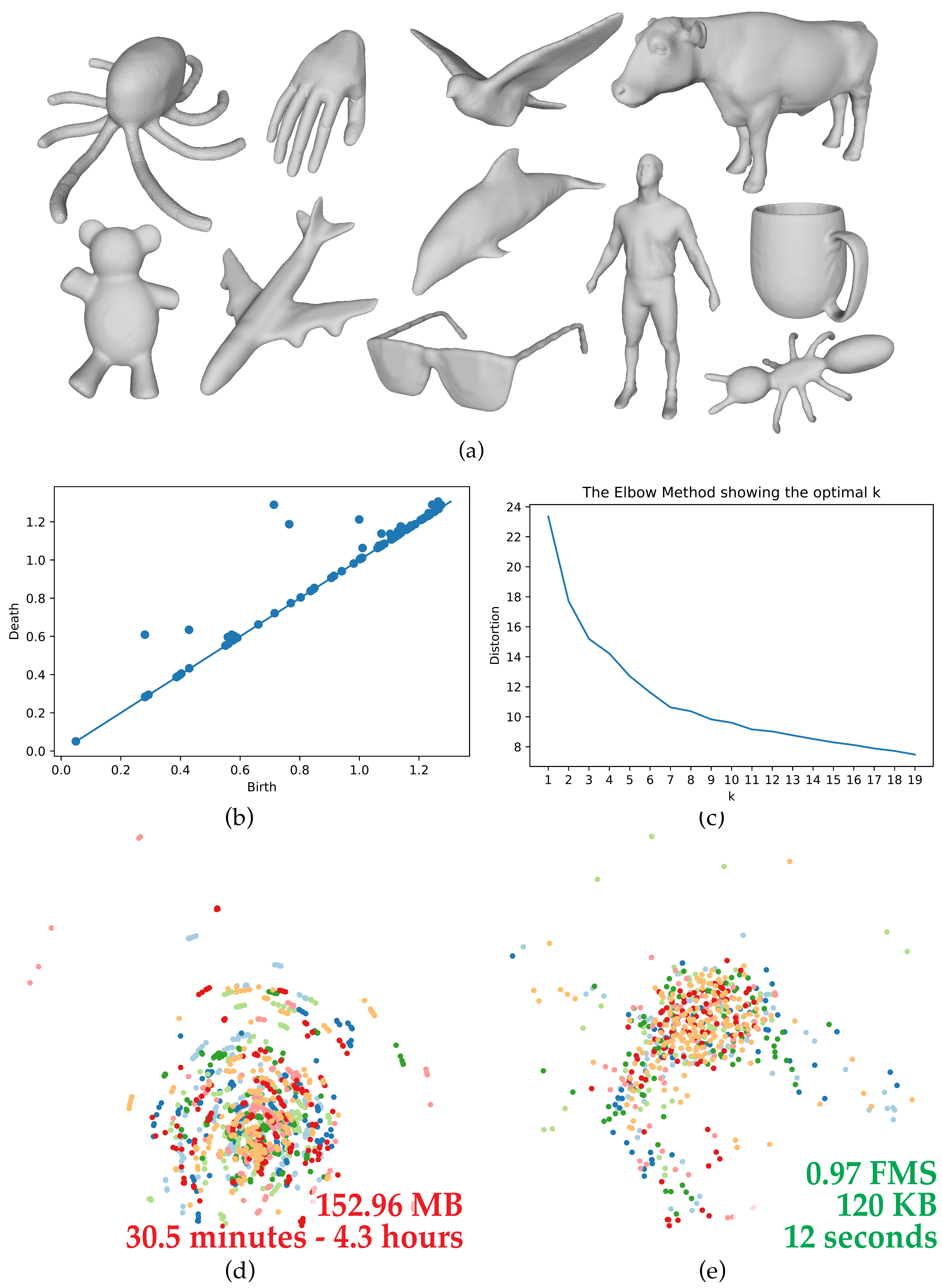}
    \caption{3D shape-2 dataset: (a) Example shape meshes from the dataset and (b) one diagram example. As this dataset does not have a known amount of topologically distinct clusters, we use (c) the elbow method to find that $k=7$ clusters exist with respect to the Wasserstein distance. (d) and (e) illustrate the MDS plots for Wasserstein distance and for Hamming distance for our generated binary codes, respectively. The FMS of 0.97 indicates that our method almost perfectly matches the original clusters.}
    \label{fig:3DShape-2}
\end{figure}

\textbf{3D Shape-2}\cite{Chen:2009:ABF}: This dataset contains 3D shapes across
19 object categories. 1,900 diagrams are produced in the same way as the
previous dataset. The average number of persistence points per diagram on this set is
22, ranging from 10 to 78. We use the elbow method \cite{thorndike1953belongs} to find that $k=7$ clusters exist. See \figref{3DShape-2}.

\begin{figure}[htb] \centering
\includegraphics[width=\columnwidth]{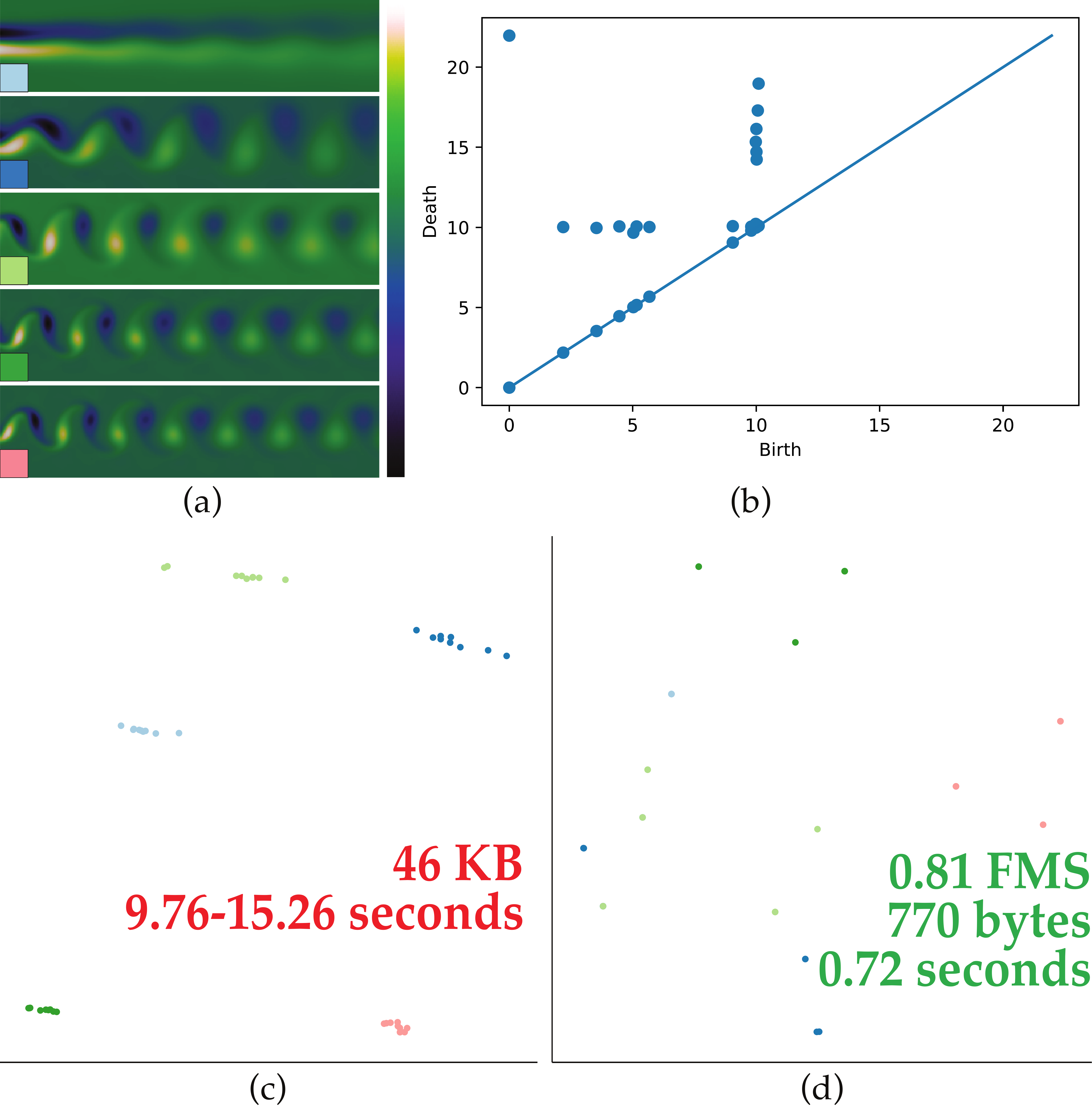} \caption{Vortex Street: This dataset contains $k=5$ clusters with (a) showing the representative data for each cluster. (b) An example diagram is provided. The MDS plot for this dataset is provided for (c) the Wasserstein distance and (d) the Hamming distance of our binary codes.}
\label{fig:vortexStreet} \end{figure}

\textbf{Vortex Street}\cite{vidal2019progressive}: This ensemble dataset
includes 45 examples of a 2D simulation of flow turbulence behind an obstacle
for $k=5$ clusters of different viscosity. The average number of persistence points in this set is 22, ranging from 20 to 50. Diagrams are produced via sublevel set filtration \cite{vidal2019progressive} using TTK~\cite{tierny2017topology}. \figref{vortexStreet} shows examples and a representative persistence diagram.

\begin{figure}[htb]
    \centering
    \includegraphics[width=\columnwidth]{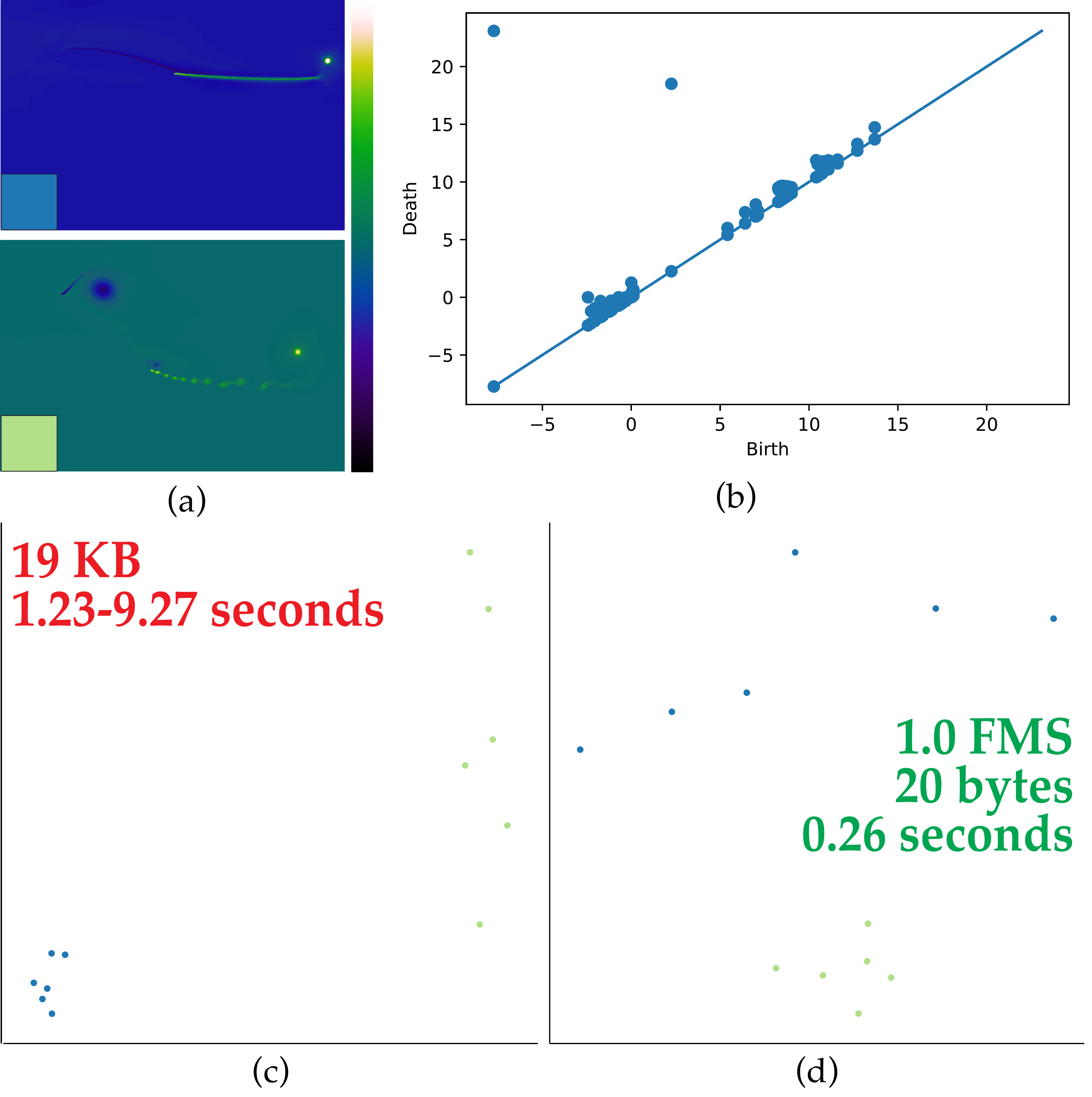}
    \caption{Starting Vortex: (a) This ensemble set contains 2 clusters. (b) An example persistence diagram is provided. (c) and (d) The FMS of 1.0 shows that our method achieves a perfect matching among clusters compared to clustering using Wasserstein.}
    \label{fig:starting_vortex}
\end{figure}

\textbf{Starting Vortex}\cite{vidal2019progressive}: This ensemble dataset
includes 12 examples of a 2D simulation with the formation of a vortex behind a
wing giving $k=2$ topological clusters. The average number of persistence points of this set of
persistence diagrams is 36 with 30 to 60 each. Diagrams are produced similarly to the previous ensemble. See \figref{starting_vortex}.

\textbf{Colorectal Cancer}\cite{kather_jakob_nikolas_2018_1214456}:  This is a
set of 10,000 regions of interest images from hematoxylin \& eosin (H\&E)
stained histological images with 9 classes. The average number of persistence points in this set is 498, ranging from 78 to 802. We conduct experiments on the full
dataset and on a subset, since other approaches cannot run on the full data. The subset contains 200 images per class and
produces 1,800 persistence diagrams in total with an average of $503$
persistence points ranging from 95 to 789. Diagrams are obtained via sublevel set filtration~\cite{lawson2019persistent} using the Giotto-tda library~\cite{tauzin2020giottotda}. We use the elbow method \cite{thorndike1953belongs} to determine that there are $k=8$ distinct topological clusters. See \figref{teaser}.

\begin{table*}[tb]
  \caption{Running times (in seconds) for the approaches outlined in this work
    to compute the distance matrix of the datasets with
    $\mathbf{N}$ diagrams with average persistence points (Avg Pts). Wasserstein
    (W1), Hera, 2D histograms (HW), progressive Wasserstein (\PW), persistence images (PI), Betti Curves (BC), and our approach are provided.  The speedup of our approach compared to the next fastest is also provided.
  }
  \label{tab:runtime-result}
  \scriptsize%
	\centering%
	\resizebox{\linewidth}{!}{%
  \renewcommand{\arraystretch}{1.4}
  \begin{tabu}{%
	lrrrrrrrrrrrrrrr%
	*{7}{c}%
	*{2}{r}%
	}
  \toprule
    & &  &W1 &Hera &HW &\PW &\multicolumn{3}{c}{PI}
   &\multicolumn{3}{c}{BC} &\multicolumn{3}{c}{Ours} & Speedup\\
  \cmidrule(lr){8-10} \cmidrule(lr){11-13} \cmidrule(lr){14-16}
Dataset &$\mathbf{N}$ &  Avg Pts  &Total &Total &Total &Total &Generate &Distance &Total &Generate &Distance &Total &Generate &Distance &Total   \\
  \midrule
  Colon Cancer &10,000 &498 &--  &-- &-- &--   &162.21 & 1208.8 & 1371.01 &76.36 & 1257.31 & 1333.67 &135.88 & 119.08 & \textbf{254.96} & 5.2X   \\

  \hline
  Colon Cancer-sub &1,800 &503 &$>$10D &$>$7D &$>$4D  &-- &31.58 & 73.04 & 104.82 &4.49 & 76.59 & 81.08 &9.36 & 3.98 & \textbf{13.34}  & 6.1X    \\

  \hline
	3D Shape-1 &1,200 &63 &78037.24 &7523.12 &1588.35  &-- &4.78 & 38.28 & 43.06   & 2.17 & 35.14 & 37.31 & 6.92 & 2.62 & \textbf{9.54} & 3.9X     \\

	\hline
  3D Shape-2 &1,900 &22  &15321.68 &1832.94 &633.52 &--  &6.48 & 67.4 & 73.88 &3.13 & 67.69 & 70.92  &8.1 & 3.9 & \textbf{12} & 5.9X     \\

  \hline
   Vortex Street &45 & 14 &15.26 &9.76 &0.63  & 0.09*  & 0.19 & 0.041 & 0.231 &0.14 & 0.04 & \textbf{0.18} &0.72 & 0.0033 & 0.72 & -    \\

   \hline
  Starting Vortex &12 &36  &9.27 &1.23  &0.14  &0.18*  &0.06 & 0.03 & 0.09 &0.02 & 0.044 & \textbf{0.064} & 0.23 & 0.026 & 0.26 & -    \\

  \bottomrule
  \end{tabu}%
  }
\end{table*}

\begin{table}[h]
  \caption{Comparison of clustering results with different training sets using
    Fowlkes-Mallows score, the Model-20 indicates we use a synthetically random
    persistence diagram with 20 persistence points each for training, 50 is with
    50 persistence points and 100 is with 100 persistence points.}
  \label{tab:fm-result-train}
  \scriptsize%
	\centering%
     \renewcommand{\arraystretch}{1.2}
	\resizebox{\linewidth}{!}{%
  \begin{tabu}{%
	lrrrrrr%
	*{7}{c}%
	*{2}{r}%
	}
  \toprule
    & & & \multicolumn{3}{c}{FMS} \\
   \cmidrule(lr){4-6}
 Dataset &$\mathbf{N}$ & Avg Pts  &Model-20 &Model-50 &Model-100 \\
  \hline
  	3D Shape-2 &1,900 &22 & \textbf{0.97}  &0.96  &0.62    \\
  	3D Shape-1 &1,200 &63 &0.77  & \textbf{0.83}  &0.67    \\
	Colon Cancer-sub &1,800 &503 & 0.71  &0.76  &\textbf{0.99}    \\
  \bottomrule
  \end{tabu}%
  }
\end{table}

\subsection{Evaluation of Clustering}
\label{sec:evaluation}
Given a set of input persistence diagrams, 
$\dgmset$, our approach computes a set
$\binarycodeset$, 
of binary representations.
To evaluate these binary representations, as well as other representations such as persistence images and Betti curves, we compare their clustering performance against clustering using Wasserstein distance.  As it is considered the standard distance for diagrams, we treat this distance as our ground truth. Let $\mathfrak{C}_1$ be the set of clusters obtained by performing hierarchical clustering on $\dgmset$ using Wasserstein. And let $\mathfrak{C}_2$ be a set of clusters obtained by performing hierarchical clustering on the set of
vectorized representations of the persistence diagrams with their associated distances (for example $\binarycodeset$ uses Hamming distance; Betti curves use Euclidean distance).

We compare these two clusterings $\mathfrak{C}_1$ and $\mathfrak{C}_2$ using the
Fowlkes-Mallows score (FMS) \cite{fowlkes1983method,m-ccibd-07} to quantify the similarity
of the clustering. This is defined as the geometric mean of precision and recall:
$$
    FMS(\mathfrak{C_1},\mathfrak{C_2})= \frac{TP}{\sqrt{{(TP+FP)(TP+FN)}}},
$$
where $TP$ (true positive) is the number of pairs of persistence diagrams that belong to the same clusters in $\mathfrak{C}_1$ and $\mathfrak{C}_2$.
$FP$ (false positive) is the number of such pairs that are in different clusters in $\mathfrak{C}_1$, but in the same cluster in $\mathfrak{C}_2$.
$FN$ (false negative) is the number of such pairs that are in the same cluster in $\mathfrak{C}_1$, but in different clusters in $\mathfrak{C}_2$. A FMS value of 1 means a perfect match with the minimum value being 0.  As our experiments show in \secref{performance}, our results are all close to 1. In addition, multidimensional scaling (MDS) \cite{borg2005modern} plots are provided in Fig. \ref{fig:3DShape-1}-\ref{fig:starting_vortex} to visualize clustering results using Wasserstein and our approach.

\subsection{Domain-Specific vs. Domain-Oblivious Training}
\label{sec:domain}
Our first approach to training is the obvious one: we train on domain-specific data to evaluate if our hash function can sufficiently preserve this space. We evaluated our approach on the 3D shape-1 dataset, splitting the 1,200 diagrams into training and test sets of size 900 and 300, respectively.  Clustering with this trained model gave a FMS of 0.83 when compared to the ground-truth clustering using Wasserstein.  While this test shows that using domain-data is a viable strategy for our approach, limiting training to domain-specific data would hinder its applicability. As with all machine learning approaches the availability of data is critical to build well-trained models. While datasets like ensemble simulations may have enough data to train the model, this is not guaranteed.  Therefore, we evaluated a more general approach.

Ideally, training should be domain-oblivious, thereby removing the need for plentiful domain data. To evaluate this possibility,
we have trained our model purely on synthetic data.  We note that persistence diagrams can be thought of as a specialized 2D
scatter plot. Therefore we can produce synthetic diagrams by creating random
scatter plots with a 
{\em uniform persistence point distribution
(rejecting points under the diagonal)}.  Training using this naive, synthetic data provided surprising results.  We trained
our model with diagrams with 50 randomly distributed persistence points and clustered 3D shape-1. The synthetic data produced the same
FMS within 0.001 of clustering using the domain-specific model. In our experiments, we found that the only requirement
for this approach is that the synthetic diagrams used in training should each have a number of points close to the average number of persistence points in the data to be clustered. This result is not only not obvious, but one would automatically think the opposite: that a set of naive, random diagrams
would not sufficiently sample the space of potential diagrams. Our experimental findings raise interesting questions
on this space that we discuss in~\secref{conclusion}.

We adopt this domain-oblivious approach as the primary method for training in this work. We train three models
for evaluation: Model-20, Model-50, and Model-100 with 4000 diagrams each with
20, 50, and 100 persistence points per diagram respectively. \tabref{fm-result-train} illustrates the requirement described above where matching the number of persistence points in training diagrams to the average number of points in the clustered data leads to higher quality results.
In our experiments, 3D shape-2
and Vortex Street were tested with Model-20. Next, 3D shape-1 and Starting Vortex were tested with Model-50. Finally, both the
full Colorectal Cancer dataset and its subset were tested using Model-100.  Note the cross-domain applicability of these models.

\subsection{Results}
\label{sec:performance}
All of our experiments were made on Intel Core 3.60GHz × 8 cores (CPU) and Nvidia GeForce GTX 1660
(the GPU was only used for training models) with 32GB of RAM. Our method is implemented
in Python with the Tensorflow platform using the implementation\footnote{\url{https://github.com/ht014/BGAN}} of Song et al.~\cite{song2018binary} for our training architecture. We also provide a lightweight C++ program for hardware-accelerated
Hamming distance computation. Our code and data are available in an OSF
repository.\footnote{\url{https://osf.io/q58c3/}}

We compare the running time and memory usage of our approach with
two popular vectorized persistence representations: Persistence
Images (PI) \cite{adams2017persistence} and Betti Curves (BC)
\cite{gameiro2004topological,robins2002computational} using GUDHI\cite{gudhi:urm}. In addition, we evaluate
our approach against two state-of-the-art approximations of Wasserstein
distance: 2D histograms with Optimal Transport (HW) \cite{lacombe2018large} and Progressive Wasserstein
(\PW)
\cite{vidal2019progressive} with their implementations. As a ground truth we compare against Wasserstein distance (W1) \cite{cohen2007stability} using scikit-tda \cite{scikittda2019} and a fast implementation of W1, the Hera method \cite{kerber2017geometry} using GUDHI \cite{gudhi:urm},
following the common parameter values for
the above. We use a PI bandwidth of $h=0.02$ with the standard weight function (1/persistence) and the entropic term for HW is $0.1/avg_{i=1}^N|p_i|$.  The grid resolution for all is $50 \times  50$ and bounded by the min/max coordinate of the diagram.  This resolution was determined through experimental evaluation of the range $[10^2,100^2]$ with a step-size of 1. We found in our testing of 3D shape-1, that sizes over 50 only provided minimal improvement of the FMS (approximately $0.001$). Therefore 50 was chosen as the minimum sized representation that still provides good quality results.  The PI bandwidth was also determined experimentally by testing the range
$[0.001,1.0]$ with step size 0.001.

\subsubsection{Speed}
\tabref{runtime-result} details our full comparison of runtimes to compute the distance matrix for all pairs in each dataset.  This is the input to single-linkage clustering and is therefore the only point at which each technique differs.
We separately list the time to {\em generate}  all representations (e.g., compute persistence images,
compute our binary code representation, etc.) and to compute the pairwise {\em
distances}. W1, Hera, and \PW have no generation time, and the  generation time for \OT is nominal
compared to the costly distance computation; therefore these times are presented
as total runtimes only.  Runtimes for \PW are provided but marked with an asterisk since direct comparison is not possible.  They use a fast C++ implementation (compared to our Python) and compute distances while clustering.  Therefore the distance calculation cannot be separated.  Parallelism was allowed for generation of the vectorizations, but distances were computed serially.  This was chosen to highlight and simplify the runtime comparisons.  As the distance computation is embarrassingly parallel, all approaches should be parallelizable with similar comparisons.

Let us first consider the runtimes for W1, Hera, and HW. As this table illustrates,
these approaches are prohibitively slow.  In fact, it was not feasible to run
them on our largest dataset with 10k diagrams. 
For a fair comparison to our Python distance calculation, we do not run \OT using GPU acceleration.
We note that Latombe et al.~\cite{lacombe2018large} showed that \OT still takes roughly $40-80$
minutes for $k$-means clustering even with GPU acceleration on $5$k persistence
diagrams with $50-100$ persistence points. Next we evaluated the \PW
approach.  This is a progressive approach, and for a fair comparison we ran it to completion. As the table illustrates, this approach is extremely fast for the
small test datasets.  Although in our testing the author's implementation of this approach quickly reached the
memory limits of our system (32 GB) and therefore did not scale to our
larger datasets (Colorectal Cancer, 3D shape-1 and 3D shape-2).  We ran \PW on a 50 image subset of the Colon Cancer dataset and it took over an hour.

Comparing to other vectorized approaches, our method offers significant performance improvements for our larger datasets giving speedups of $3.9X\sim6.1X$ when compared to the fastest alternative approach.  As the table shows, our distance computation is incredibly fast leaving the bottleneck of our approach to be the generation of the binary codes.  As such, for datasets that are small, where distance computation does not dominate, our runtimes are comparable to other vectorized approaches.

As a final illustration of the speed, we have implemented the distance calculation of our approach in C++ and leverage the population count and XOR support that exists on modern CPUs. These results are provided for our largest datasets in \tabref{c-result}. The runtimes are 2-3 orders of magnitude faster than our Python implementation and take all-pairs distance computation down to just milliseconds. Overall, these results show the speed and scalability of our proposed method and
how well positioned our binary codes would be for large-scale tasks or databases.
\begin{table}[h]
  \caption{Timings for comparisons using a C++ implementation that leverages hardware acceleration to compute Hamming distances.}
  \label{tab:c-result}
  \scriptsize%
  \centering%
  \renewcommand{\arraystretch}{1.2}
	\resizebox{0.8\linewidth}{!}{%
  \begin{tabu}{%
	lrrrr%
	*{7}{c}%
	*{2}{r}%
	}
  \toprule
   Dataset  &$\mathbf{N}$  & Python & C++ & Speedup \\
   \hline
   Colon Cancer &10,000 & 119.08s  &161.61ms  &737X \\
	3D Shape-1 &1,200 & 2.62s  &2.27ms &1154X  \\
  3D Shape-2 &1,900   & 3.9s  &5.32ms  &733X \\
  \bottomrule
  \end{tabu}%
 }
\end{table}

\subsubsection{Memory and Storage}
\tabref{size-result} describes the potential memory and storage gains from our
representation.  The 2D persistence points of the diagrams (PD) were saved using 64-bit precision.  The grids for PI/BC were also saved with 64-bit precision.  We compare this requirement to our approach that saves a single 64-bit number. Reduction rate is compared to the next smallest representation. As this table illustrates, as one would expect, saving a single integer can have significant benefits for storage and memory. The encoder size for our approach, which would also have to be saved, is not included in these results. This cost can vary.  Our encoder for this work was the same as Song et al.~\cite{song2018binary}: a standard, pretrained VGG19~\cite{szegedy2015going} model, which is approximately 500MB. However, it may be possible to compress this size~\cite{cheng2017survey} . For instance, work has shown VGG19 can reduce its parameters  $5-20X$~\cite{masana2017domain}. Other models like SqueezeNet~\cite{iandola2016squeezenet} reduce to less than 0.5MB. Overall, since our approach is domain-oblivious, a single model can be potentially used on a multitude of datasets from a large number of domains. Thus the cost of storing the encoder would be amortized in practice and will not grow as data sizes increase. Therefore this approach has the potential to greatly reduce storage overheads as the use of TDA grows.
\begin{table}[h]
  \caption{Comparison of size (in MB) with persistence summaries. As vectorized
    summaries of persistence diagrams, here PI and BC have the same vector size ($50x50$).}
  \label{tab:size-result}
  \scriptsize%
	\centering%
  \renewcommand{\arraystretch}{1.4}
	\resizebox{\linewidth}{!}{%
  \begin{tabu}{%
	lrrrrr%
	*{7}{c}%
	*{2}{r}%
	}
  \toprule
   Dataset  &$\mathbf{N}$ & Avg Pts  &PD &PI/BC & Ours (64bits) & Reduction \\
   \hline
   Colon Cancer &10,000 &498  &800.96   &201.21  &\textbf{0.64} &  192X  \\
  Colon Cancer-sub &1,800 &503  &144.96   &36.2  &\textbf{0.12}  & 254X \\
	3D Shape-1 &1,200 &63 &130.12   &24.13 &\textbf{0.08} & 301X  \\
  3D Shape-2 &1,900 &22   &152.96  &36.2 &\textbf{0.12} &  301X   \\
  Vortex Street &45 & 14  &0.046   &0.9 &\textbf{0.00077} & 1168X    \\
  Starting Vortex  &12 &36  &0.019   &0.24 &\textbf{0.00002} & 1200X  \\

  \bottomrule
  \end{tabu}%
 }
\end{table}

\subsubsection{Quality}
\tabref{fm-result} shows the evaluation of 
clustering quality, comparing our
clustering results with other methods using the evaluation method described in \secref{evaluation}.  The quality is determined by the
FMS between the clusterings obtained for the different persistence diagram representations compared to the clustering
produced when using
Wasserstein distance directly on the persistence diagrams.  As such, we could only run comparisons on a
1,800 diagram subset of the colorectal cancer dataset since running the full dataset was
not possible (W1, Hera) or failed (\PW) for some approaches.  For our approach, we provide the FMS
for our two methods for forming a similarity matrix: real-valued
or binary similarity.  As this table illustrates, our approach provides
comparable or better quality results when compared to progressive Wasserstein (\PW), persistence images (PI), or Betti Curves (BC).
We time-limit \PW to the total runtime of our approach as reported in \tabref{runtime-result}.
\PW would, in time, converge to the exact Wasserstein distance. Therefore for a fair comparison, we only look at their quality
for the same amount of running time.  Not only are the results from the binary codes on par with other approaches, our approach
almost achieves perfect reproduction of the clustering for the Colorectal Cancer and 3D Shape-2 datasets.  Moreover, it perfectly
matches the clustering of Starting Vortex.  Note these results use our domain-oblivious training approach and therefore the same
models were applied to more than one type of data.

\begin{table}[ht]
  \caption{Comparison of clustering results using the Fowlkes-Mallows score, as described in \secref{evaluation}.  Scores range from 0 to 1; a score of 1 indicates identical clusters. The clusterings using different persistence diagram representations (and their distances) are compared to the Wasserstein-based clustering of the input persistence diagrams.}
  \label{tab:fm-result}
  \scriptsize%
	\centering%
  \renewcommand{\arraystretch}{1.2}
  \begin{tabu}{%
	lrrrrrr%
	*{7}{c}%
	*{2}{r}%
	}
  \toprule
    & & & & & \multicolumn{3}{c}{Ours} \\
   \cmidrule(lr){6-8}
 Dataset & Avg Pts &\PW  &PI &BC &Model &Real &Bin \\
  \hline
  Colon-sub &503  &0.71    &0.98  &\textbf{0.99}& 100 &0.92 &\textbf{0.99}   \\
	3D Shape-1 &63 &0.54   & 0.8 &0.81 & 50 &  0.82 &\textbf{0.83} \\
  3D Shape-2 &22   &0.74  &\textbf{0.97} &0.96 & 20 & 0.91 &\textbf{0.97}  \\
     Vortex Street & 14  &\textbf{1}   &0.79  &0.78  & 20 &  0.80 & 0.81  \\
  	  Starting Vortex &36  &\textbf{1}   &\textbf{1} &\textbf{1}  & 50 & 0.63 &\textbf{1}\\
  \bottomrule
  \end{tabu}%
\end{table}

\begin{table}[th]
  \caption{Comparison of clustering results with different number of bits, using the Fowlkes-Mallows score.}
  \label{tab:fm-result-bits}
  \scriptsize%
	\centering%
    \renewcommand{\arraystretch}{1.2}
    \resizebox{\linewidth}{!}{%
  \begin{tabu}{%
	lrrrrrr%
	*{7}{c}%
	*{2}{r}%
	}
  \toprule
 Dataset &$\mathbf{N}$ & Avg Pts  &24 bits &48 bits & 64 bits &128 bits &256 bits \\
  \hline
  	3D Shape-1 &1,200 &63 &0.64  &0.75  &\textbf{0.83} &0.84  &0.86 \\
  	  	3D Shape-2 &1,900 &22 &0.82 &0.89 &\textbf{0.97} &0.97 &0.98 \\
  	Vortex Street &45 &14 &0.67 &0.77 &\textbf{0.81} &0.83 &0.86 \\
  	  	Starting Vortex &12 &36 &0.94 &1 &\textbf{1} &1 &1 \\
  \bottomrule
  \end{tabu}%
 }
\end{table}

\begin{figure}
    \centering
         \includegraphics[width=\columnwidth]{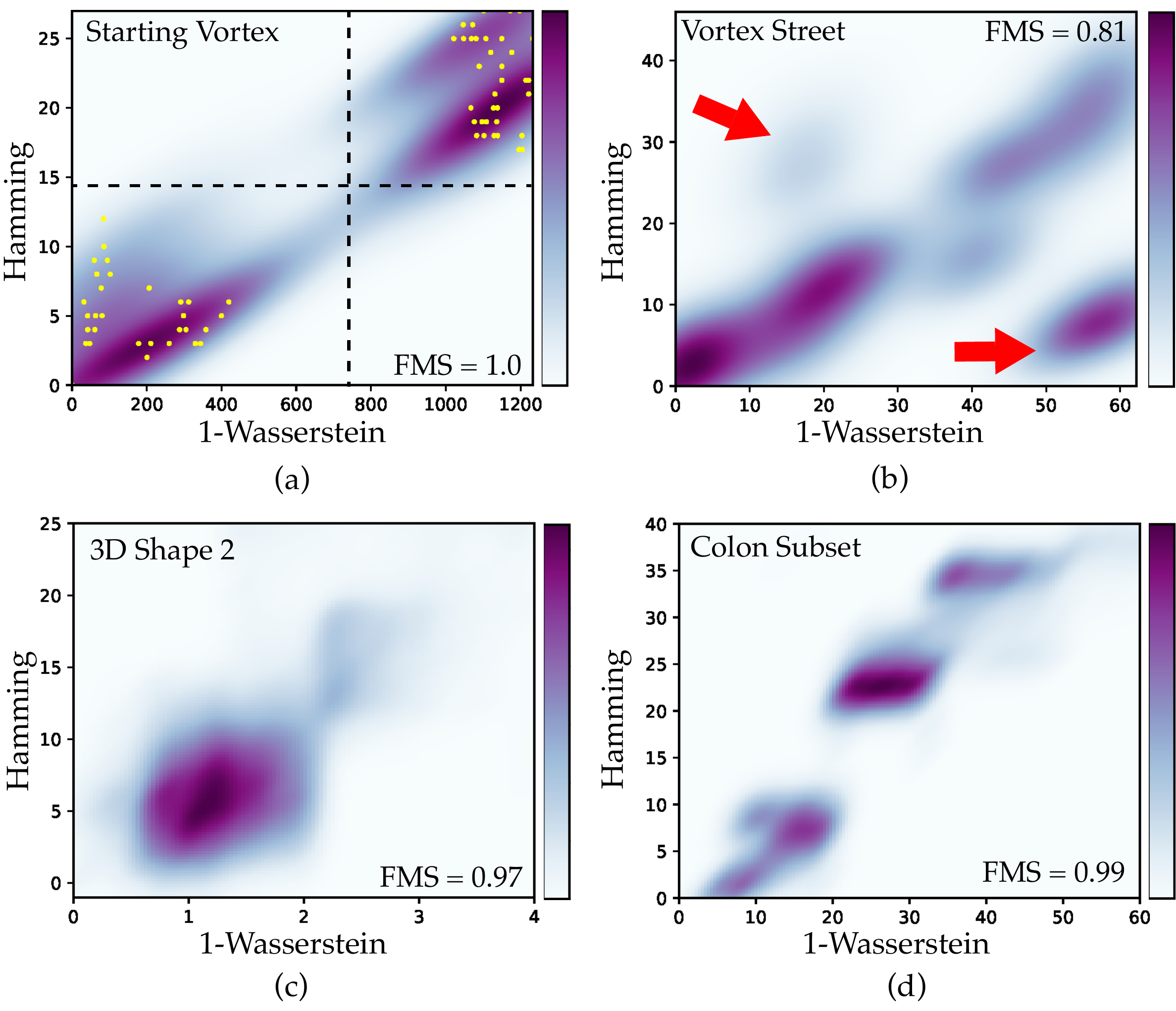}
        \caption{ (a-d) Plots to illustrate distance preservation using binary codes for various datasets. Each is a scatterplot of points for each diagram whose horizontal position is W1 distance and vertical is Hamming. Each is drawn with a kernel density estimator.  All exhibit a linear, diagonal shape meaning that distances are being preserved.  (a) also plots the points directly (yellow). There is a clear separation of the two clusters both horizontally and vertically.
        (b) highlights the two areas that break this linear property that likely give its lower FMS.
         }
        \label{fig:scatterplot}
\end{figure}

To further evaluate the quality of distance preservation using our binary codes, we provide scatterplots in Fig. \ref{fig:scatterplot} by setting Wasserstein and Hamming distance as x-y coordinates for each point pair.  To avoid overdrawing, each point is drawn as a Gaussian kernel density estimator.  Each point in the plot is a diagram with the horizontal position given by the W1 distance and vertical being the Hamming distance.  If distances are being maintained, this plot should be linear about a diagonal.  As this figure illustrates, this is the case for our binary codes. In \figref{scatterplot} (a) the points are provided as well (yellow). This shows there is clear separation in both dimensions (dotted line) meaning that these clusters are well-maintained in both distances.  Finally, \figref{scatterplot} (b) illustrates the plot for Vortex Street, an example with a lower FMS.  As highlighted with red arrows, there are clear clusters of points where distance is not being maintained and likely give the lower score (although still higher quality than PI and BC).

Next, we evaluate quality in FMS in relation to the number of bits used in the binary code.  We experimented with the clustering result of 3D shape-1, 3D shape-2, Vortex Street, and Starting Vortex  by varying bit lengths from 24-256.  The results of this experiment are provided in \tabref{fm-result-bits}.  As this table shows, and as one would expect, increases in bit length result in increases in quality of the final result, although with a noticeable falloff in gains after 64 bits. Therefore we opted for 64-bit codes since they provide high-quality results with simpler implementations than the higher bit counts.

\tabref{fm-result-similarity} illustrates a FMS comparison of clustering the 3D Shape-1 dataset with different similarity matrix strategies. In this table, Real denotes the real-valued similarity matrix. S-X denotes the use of a binary similarity matrix built from the real-valued distance matrix. S-1 uses fixed number of k-nearest neighbors, where $k=1000$ out of the $4000$ training set. Similarly, S-2 uses $k=600$ and S-3 uses $k=1400$. For a soft threshold similarity matrix, S-4 uses a strategy of limiting similarity to use a global threshold of 25\% percent. Finally, S-5 is our two pass mean rejection.  As this figure illustrates, the two pass approach leads to more accurate clustering and is therefore used by our work.

\begin{table}[h]
  \caption{Comparison of clustering results with different similarity matrix computation methods, using the Fowlkes-Mallows score.}
  \label{tab:fm-result-similarity}
  \scriptsize%
	\centering%
     \renewcommand{\arraystretch}{1.2}
	\resizebox{\linewidth}{!}{%
  \begin{tabu}{%
	lrrrrrr%
	*{7}{c}%
	*{2}{r}%
	}
  \toprule
 Trained Model &$\mathbf{N}$ & Avg Pts  &Real &S-1 &S-2 &S-3 &S-4 &S-5 \\
  \hline
  	Model-50 &4,000 &50 &0.82  &0.77  &0.78 &0.8 &0.81 &\textbf{0.83}   \\
  \bottomrule
  \end{tabu}%
}
\end{table}

\section{Conclusions}
\label{sec:conclusion}

In the paper, we present an approach to produce concise binary codes of
persistence diagrams that maintain topological similarity.
The key to this approach is the training of a machine learning model that learns a hash,
not on domain-specific data, but on randomly generated 2D scatter plots.  This
leads to a technique that is domain-oblivious, where a model can be applied
across multiple domains or types of data without the need for retraining. As
this is a hashing approach, our technique is not likely to maintain small distances.
For applications where close distances are discounted, our
approach is well-suited.  It is still an open question if a hashing approach
could be designed such that small distances are maintained.
The data used in our synthetically trained model only needs to roughly match the
average number of persistence points of the testing dataset.  In practice, we have found this is not an overly strict requirement. For instance, the Colorectral Cancer dataset has
500 persistence points on average, but Model-100 worked well in our tests. In regards to
storage, while our binary code is small, one would still need to save the
encoder, which for deep networks can be many MB.  As we mentioned,
given that a single model can be applied to many datasets across many domains, we argue that the amortized cost could be 
nominal in practice. Although, we plan to explore how to reduce this overhead in future work. Finally, this work illustrated the benefits
of this representation through examples from topological clustering, where our new binary codes
provide fast, high-quality results.  Moreover, the scalability
of such an approach was highlighted through the potential low storage requirements of the binary codes along
with extremely fast distance computations using on-chip acceleration.

\acknowledgments{This work was supported by the National Science Foundation and
the National Institutes of Health (NSF DMS 1664848 and 1664858), NSF CCF 2046730, and NSF IIS 2136744.}

\bibliographystyle{abbrv-doi}

\bibliography{main}
\end{document}